\documentclass[11pt,a4paper]{article}

\usepackage{amsfonts}
\usepackage{color}
\usepackage{graphicx}

\title{\textbf{Asymmetric kink scattering in a two-component scalar field theory model}}

\author{A. Alonso-Izquierdo$^{(a)}$
\\ {\normalsize {\it $^{(a)}$ Departamento de Matematica
Aplicada}, {\it Universidad de Salamanca, SPAIN}} }

\date{}

\begin{document}

\maketitle

\begin{abstract}

In this paper the kink scattering in a two-component scalar field theory model in (1+1)-Minkowskian space-time is addressed. The potential term $U(\phi_1,\phi_2)$ is given by a polynomial of fourth degree in the first field component and of sixth degree in the second one. The novel characteristic of this model is that the kink variety describes two different types of extended particles. These particles are characterized by its topological charge but also by a new feature determined by a discrete charge $\Lambda=0,\pm 1$. For this reason, the kink scattering involves a very rich variety of processes, which comprises kink annihilation, reflection, charge exchange, transmutation, etc. It has been found that not only the final velocity of the scattered kinks, but also the final nature of the emerging lumps after the collision are very sensitive on the initial velocities. Asymmetric scattering processes arise when Type I and Type II particles are obliged to collide. In this case, ten different final scenarios are possible. Symmetric scattering events are also discussed.   
\end{abstract}

\section{Introduction}

Topological defects are solutions in field theory models which cannot decay to the vacuum because of topological constraints in the configuration space. For some physical systems they can be interpreted as extended particles because its energy density is localized. The characteristics of these solutions have been exploited in several disciplines in order to explain new phenomena in non-linear sciences. To mention some examples, topological defects have been applied in Condensed Matter Physics to explain the behavior of ferroelectic materials \cite{Eschenfelder1981,Jona1993}, in Cosmology to understand the Early Universe \cite{Vilenkin1994, Vachaspati2006,Gani2018b}, in Optics to describe some properties of signal transmission in optical fibers \cite{Mollenauer2006, Schneider2004, Agrawall1995}, in Biochemistry to clarify some features of DNA \cite{Yakushevich2004} and other substances \cite{Davydov1985}, etc. As a consequence, the scattering between topological defects has received much attention and has been extensively studied both in Physics and Mathematics.

In the case of scalar field theory models, such as those considered in this paper, this type of solutions (referred to as \textit{kinks}) must comply with non-linear Klein-Gordon partial differential equations, which are, in general, non-integrable systems. Curiously, kink scattering is more complex in these cases than in integrable systems. In fact, the study of this issue has led to the discovery of very interesting and unexpected properties. For example, the $\phi^4$-model involves the presence of two vacua. The kink variety in this case comprises two topological defect solutions joining these points, the \textit{kink} and the \textit{antikink}, which carry opposite topological charge. The only possible scattering event in this model is given by the kink-antikink collision. This scattering process has been studied in the seminal papers \cite{Sugiyama1979,Campbell1983,Anninos1991,Takyi2016}. There are only two types of final scenarios, whose presence depends critically on the initial velocity: (a) \textit{bion formation} (kink and antikink collide and bounce indefinitely radiating energy in every impact) and (b) \textit{kink reflection} (where the kink and antikink eventually are able to escape with some final separation velocity $v_f$). One of the most remarkable aspects of this model is that the transition between the previous regimes involves the presence of resonance windows with a fractal structure where the previous regimes are interlaced and the kinks must collide a finite number of times before definitely escaping. This behavior is explained by the so called \textit{resonant energy transfer mechanism} where an energy exchange takes place between the zero and vibrational kink modes, see \cite{Campbell1983}. An analytical explanation of this phenomenon is given in References \cite{Goodman2005,Goodman2008,Goodman2007}. The kink scattering together with the presence of the resonance windows have been explored in other models, such as in the double sine-Gordon model \cite{Shiefman1979,Peyrard1983,Campbell1986,Gani1999, Malomed1989,Gani2018}, in deformed $\phi^4$ models \cite{Simas2016,Gomes2018,Bazeia2017b,Bazeia2017a}, in $\phi^6$-models \cite{Dorey2011,Weigel2014,Gani2014,Bazeia2018b,Romanczukiewicz2017}, in $\phi^8$-models \cite{Belendryasova2017,Christov2018,Gani2015}, in models with defects, impurities or inhomogeneities \cite{Fei1992,Fei1992b,Malomed1992,Goodman2004,Javidan2006,Saadatmand2012,Saadatmand2015,Saadatmand2018, Adam2018}, for the coupled nonlinear Schrodinger equations with vector solitons \cite{Yang2000,Tan2001,Goodman2005b}, etc. In these cases the resonant energy transfer mechanism is activated by the presence of vibrational modes of the single kinks or of a combined kink-antikink configuration. A review of recent works on this issue is given in Reference \cite{Romanczukiewicz2018}.
The role of quasi-normal modes in the existence of this phenomenon has also been investigated, see \cite{Dorey2018}. The collision of $N$ kinks has been recently studied in \cite{Marjaneh2017,Marjaneh2017b,Marjaneh2017c} for different models. A new area of research nowadays corresponds to the study of the dynamics of kinks with power-law asymptotics, see \cite{Christov2018, Gomes2012, Radomskiy2017, Bazeia2018c, Bazeia2018d, Christov2018b,Manton2019}.

In one-component scalar field theory models all the possible scattering events obtained from an initial two-kink configuration with zero topological charge reduce to the collision between a kink and its own antikink. This situation is dramatically changed for models with two or more scalar fields. Kinks joining the same vacuum points can follow distinct orbits in the internal plane and, therefore they describe different types of extended particles. The analytical identification of kink solutions for these models is a difficult task, which has led to an active research area during the last decades. For example, exact kink solutions have been obtained for the MSTB model \cite{Ito1985,Alonso1998}, the generalized MSTB models \cite{Alonso2008b} and its extensions to three fields \cite{Alonso2000,Alonso2002c,Alonso2004}, the BNRT model \cite{Bazeia1995,Shifman1998,Alonso2002}, nonlinear massive Sigma models \cite{Alonso2008,Alonso2009,Alonso2010,Alonso2018c}, coupled $\phi^4$ and sine-Gordon models \cite{Katsura2014}, models with a real scalar Higgs field and a scalar triplet field \cite{Gani2016}, etc. Some deformation procedures have been developed to obtain exact solutions of two-field models from one-field models, see \cite{Bazeia2013}. Domain walls coupled to fermionic degrees of freedom have been studied in \cite{Gani2010,Gani2011}. The usual practice in studying kink dynamics in this type of models has been to consider some adiabatic approximations, see \cite{Morris2018, Alonso2002d,Alonso2006}, which are valid for kink collisions with low initial velocities. More general analysis of kink dynamics in two-component scalar field theory models has been carried out in recent works, see for instance \cite{Romanczukiewicz2008, Halavanau2012, Alonso2017, Alonso2018, Alonso2018b, Ferreira2019}.

In this paper we are interested in the study of the kink scattering in a particular two-component scalar field theory model in (1+1)-Minkowskian space-time with a potential term $U(\phi_1,\phi_2)$ given by a polynomial of fourth degree in the first field and sixth degree in the second one. The novel characteristic of this model is that the kink variety describes two different types of extended particles with different energy density distribution. These particles are characterized by its topological charge but also by a new feature described by a discrete charge $\Lambda=0,\pm 1$, which can be modified by the kink collision giving rise to a new pair of emerging lumps. For this reason, the kink scattering in this case involves a very rich variety of processes, which comprises kink annihilation, reflection, charge exchange, transmutation, etc. It has been found that not only the final velocity $v_f$ of the scattered kinks, but also the final nature of the emerging lumps after the collision are very sensitive on the initial velocities. Asymmetric scattering processes arise when Type I and Type II particles are obliged to collide. In this case, ten different final scenarios can be found.  

The organization of this paper is as follows: in Section 2 the model is introduced and the topological kinks describing the Type I and Type II particles are analytically identified. The linear stability study of these solutions is also addressed. Section 3 is devoted to the analysis of the kink scattering of these particles. Firstly, a general discussion of the kink scattering processes in this case is portrayed. The asymmetric scattering events given by the collision between Type I and Type II particles are described in Section 3.1 whereas symmetric events are considered in Section 3.2 and Section 3.3. Finally, some conclusions are drawn in Section 4.

\section{The model and its static kink variety}

We shall deal with a (1+1)-dimensional two-coupled scalar field theory model whose dynamics is governed by the action
\begin{equation}
S=\int d^2 x \Big[ \frac{1}{2} \partial_\mu \phi_a \partial^\mu \phi_a - U(\phi_1,\phi_2) \Big] \hspace{0.4cm},
\label{action}
\end{equation}
where Einstein summation convention is assumed for $\mu=0,1$ and $a=1,2$. The model involves two dimensionless real fields $\phi_a: \mathbb{R}^{1,1} \rightarrow \mathbb{R}$ ($a=1,2$). The Minkowski metric $g_{\mu \nu}$ has been chosen as $g_{00}=-g_{11}=1$ and $g_{12}=g_{21}=0$. The spacetime coordinates will be denoted as $x^0\equiv t$ and $x^1\equiv x$ from now on. In this paper we shall explore the properties of the kink solutions for the potential term
\begin{equation}
U(\phi_1,\phi_2)=\frac{1}{2} \phi_2^2 (\phi_1^2+\tau^2 \phi_2^2 -1)^2 + \frac{1}{2} \tau^2 \beta^2 (\phi_1^2+\phi_2^2-1)^2 + \frac{1}{2} (\tau^2-1)\beta^2 \phi_2^2 \hspace{0.3cm}.  \label{potential}
\end{equation}
$U(\phi_1,\phi_2)$ is a polynomial function of fourth degree in the first field component $\phi_1$ and of sixth degree in the second component $\phi_2$, which involves the real coupling constants $\tau$ and $\beta$, i.e., $\tau,\beta\in \mathbb{R}$. Therefore, the expression (\ref{potential}) characterizes a two-parameter family of models, whose members are labeled by the points $(\tau,\beta)$ in the parameter space. Not all the members of this two-component field theory family involve the presence of topological defects. As a first requirement, the potential $U(\phi_1,\phi_2)$ must be non-negative. This leads to the condition
\begin{equation}
\tau> 1 \hspace{0.3cm} . \label{condi1}
\end{equation}
Under this assumption, the set ${\cal M}$ of vacua (absolute minima of the potential $U(\phi_1,\phi_2)$) consists of two elements
\begin{equation}
{\cal M} = \{ A_\pm =(\pm 1,0) \} \hspace{0.5cm} . \label{vacua}
\end{equation}
These values correspond to the simplest solutions of the model (zero energy static homogeneous solutions), which are fixed points of the coupled non-linear Klein-Gordon equations
\begin{eqnarray}
\frac{\partial^2 \phi_1}{\partial t^2} - \frac{\partial^2 \phi_1}{\partial x^2} &=& -2 \phi_1 \Big[ \phi_2^2 (\phi_1^2 + \tau^2 \phi_2^2-1 )+ \tau^2 \beta^2 (\phi_1^2+\phi_2^2-1) \Big] \hspace{0.5cm} , \label{kleinequations}\\
\frac{\partial^2 \phi_2}{\partial t^2} - \frac{\partial^2 \phi_2}{\partial x^2} &=&  -\phi_2 \Big[ 2 \tau^2 \phi_2^2(\phi_1^2+\tau^2\phi_2^2-1) + (\phi_1^2+\tau^2 \phi_2^2-1)^2 + 2 \tau^2 \beta^2 (\phi_1^2+\phi_2^2-1) + \beta^2 \overline{\tau}^2 \Big]\hspace{0.3cm} , \nonumber
\end{eqnarray}
derived from the Euler-Lagrange equations of the functional (\ref{action}). For the sake of simplicity, the notation $\overline{\tau}^2=\tau^2-1$ will be used in the subsequent expressions. The second order small fluctuation operator valued on the points $A_\pm$,
\[
{\cal H}[A_\pm]=\left( \begin{array}{cc} -\frac{d^2}{dx^2} + 4 \tau^2 \beta^2 & 0 \\ 0 & -\frac{d^2}{dx^2} +\overline{\tau}^2 \beta^2 \end{array} \right)
\]
provides us with insight in the linear stability of the vacua. Two continuous spectra emerge on the threshold values $4 \tau^2 \beta^2$ and $\overline{\tau}^2 \beta^2$ in this case. Thus, all the eigenvalues in the spectrum of ${\cal H}[A_\pm]$ are positive, which guarantees that the vacua $A_\pm$ are stable solutions.

On the other hand, although the origin of the internal plane $(\phi_1,\phi_2)=(0,0)$ is a solution of the system of partial differential equations (\ref{kleinequations}), it is an unstable solution. There always exist negative eigenvalues in the spectrum of the first component of the Hessian operator for this point
\begin{equation}
{\cal H}[(0,0)]=\left( \begin{array}{cc} -\frac{d^2}{dx^2} -2 \tau^2 \beta^2 & 0 \\ 0 & -\frac{d^2}{dx^2} +1-\beta^2-\tau^2 \beta^2 \end{array} \right) \hspace{0.3cm} . \label{originhess}
\end{equation}
In this paper we are interested in investigating the scattering processes between asymmetric types of stable kinks. As discussed later this scheme will be attained if the following condition on the coupling constants
\begin{equation}
1-\beta^2-\tau^2 \beta^2>0 \label{condi2}
\end{equation}
is imposed. This implies that the constant potential well of the second component in (\ref{originhess}) is positive, which turns the origin of the internal plane into a saddle point of the potential term $U(\phi_1,\phi_2)$, see Figure 1 (left). This situation has not been previously explored and it gives rise to the presence of three different stable topological kinks belonging to the same topological sector. The analysis of the scattering between these topological defects is the main goal in this paper. In Figure 1 (left) the potential function $U(\phi_1,\phi_2)$ has been depicted for the values $\tau=1.2$ and $\beta=0.2$. The restriction of $U(\phi_1,\phi_2)$ to the axis $\phi_1$ leads to the expression $U(\phi_1,0)=\frac{1}{2} \tau^2\beta^2 (1-\phi_1^2)^2$. This implies that a $\phi_1^4$-model is immersed in our scalar field theory model and a $\phi_1^4$-type kink joining the vacua $A_\pm$ will arise on the $\phi_1$-axis. The restriction of the potential function to the $\phi_2$-axis leads to the sixth order polynomial $U(0,\phi_2)=\frac{1}{2} \phi_2^2 (1-\tau^2\phi_2^2)^2 + \frac{1}{2} \tau^2\beta^2 (1-\phi_2^2)^2 + \frac{1}{2} (\tau^2-1)\beta^2 \phi_2^2$. This function has three minima (one of them located at the origin) and two local maxima. There is also room for new kink solutions joining the vacua $A_\pm$, which must be confined between the previous maxima and the potential wall, see Figure 1 (left). All these solutions are analytically identified below. In Figure 1 (right) the difference $\Delta U$ between the value of the potential $U(\phi_1,\phi_2)$ at the local maxima and at the origin as a function of the coupling constant $\beta$ is graphically represented for several values of the parameter $\tau$. This magnitude plays an important role in the stability of the previously mentioned $\phi^4$-type kinks when they are perturbed. The greater this magnitude is the more stable these solutions are when non-small fluctuations are applied. From the behavior of $\Delta U$, it is expected these kink solutions to be more stable for values of $\tau$ close to 1, see Figure 1 (right).

\begin{figure}[h]
\centerline{\includegraphics[height=3.5cm]{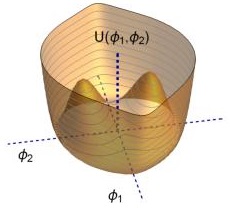} \hspace{1cm}
\includegraphics[height=3.5cm]{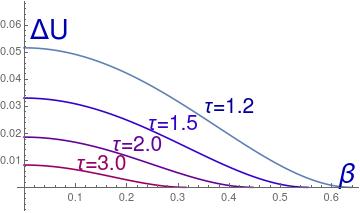} }
\caption{\small Graphics of the potential term $U(\phi_1,\phi_2)$ for the parameter values $\tau=1.2$ and $\beta=0.2$ (left). Notice that this case complies with the condition (\ref{condi2}) and the origin is a saddle point. Potential jump $\Delta U$ as a function of the coupling constant $\beta$ for several values of the parameter $\tau$ (right).}
\end{figure}

The action functional (\ref{action}) is invariant by the symmetry group $\mathbb{G}=\mathbb{Z}_2\times \mathbb{Z}_2$ generated by the transformations $\pi_1:(\phi_1,\phi_2)\mapsto (-\phi_1,\phi_2)$ and $\pi_2:(\phi_1,\phi_2)\mapsto (\phi_1,-\phi_2)$. The mirror reflection in the space coordinate $\pi_x: x \mapsto -x$ is also a symmetry. On the other hand, the spacetime translational symmetry, which arises in this type of scalar field theories, guarantees the conservation of the total energy \begin{equation}
E[\Phi(x,t)]= \int_{-\infty}^\infty dx \,\, \varepsilon[\Phi(x,t)] \hspace{0.4cm}, \label{totalenergy}
\end{equation}
for solutions $\Phi(x,t)=(\phi_1(x,t),\phi_2(x,t))$ of the field equations (\ref{kleinequations}). This quantity has been expressed in (\ref{totalenergy}) as the integral over the space $\mathbb{R}$ of the energy density
\begin{equation}
\varepsilon[\Phi(x,t)]= \frac{1}{2} \Big( \frac{\partial \phi_1}{\partial t} \Big)^2  + \frac{1}{2} \Big( \frac{\partial \phi_2}{\partial t} \Big)^2 +\frac{1}{2} \Big( \frac{\partial \phi_1}{\partial x} \Big)^2  +\frac{1}{2} \Big( \frac{\partial \phi_2}{\partial x} \Big)^2 + U(\phi_1(x,t),\phi_2(x,t)) \hspace{0.3cm}.
\label{energydensity}
\end{equation}
Besides, the configuration space ${\cal C}$ for this type of systems is restricted to the set of maps $\Phi :\mathbb{R}^{1,1} \rightarrow \mathbb{R}\times \mathbb{R}$, whose total energy is finite, i.e., ${\cal C}=\{\Phi(x,t) \in \mathbb{R}\times \mathbb{R} : E[\Phi(x,t)]<+\infty \}$. All the elements of ${\cal C}$ satisfy the following asymptotic conditions
\begin{equation}
\lim_{x\rightarrow \pm \infty} \frac{\partial \Phi(x,t)}{\partial t} = \lim_{x\rightarrow \pm \infty} \frac{\partial \Phi(x,t)}{\partial x} = 0 \hspace{0.5cm},\hspace{0.5cm} \lim_{x\rightarrow \pm \infty} \Phi(x,t)\in {\cal M}  \hspace{0.3cm}. \label{asymptotic}
\end{equation}
Therefore, the configuration space ${\cal C}$ is the union of four topologically disconnected sectors, ${\cal C}=\cup_{i,j=1}^2 {\cal C}_{ij}$. Every sector is characterized by the asymptotically connected elements in ${\cal M}$, as pointed out by the relation (\ref{asymptotic}). As a consequence, the topological charge
\[
q =\frac{1}{2} \Big( \phi_1(+\infty,t)-\phi_1(-\infty,t) \Big)
\]
is an invariant of the system. The topological defect solutions of (\ref{kleinequations}) carry non-zero topological charge $q$. In general, a topological defect with a positive topological charge will be referred to as \textit{kink} whereas the term \textit{antikink} will be used to name solutions with negative topological charge.

Now, the static kinks (time-independent finite energy solutions of the field equations (\ref{kleinequations}) whose energy density (\ref{energydensity}) is localized) will be analytically identified. There exist two types of static kinks in the model, both of them joining the vacuum points $A_\pm$:

\vspace{0.2cm}

\noindent (I) \textsc{One-component topological kinks:} If the trial orbit $\phi_2=0$ is plugged into the partial differential equations (\ref{kleinequations}), the static topological kink
\begin{equation}
K^{(q,0)}_{\rm static}(\overline{x}) =\Big( q \, \tanh (\beta \tau \overline{x}) ,0 \Big) \label{tk1}
\end{equation}
can be easily identified. Here, $\overline{x}=x-x_0$ where $x_0\in \mathbb{R}$ fixes the kink center and $q=\pm 1$ is the topological charge of the solution. In particular, the kink $K^{(1,0)}_{\rm static}(\overline{x})$ asymptotically goes from the vacuum $A_-$ at $x=-\infty$ to the vacuum $A_+$ at $x=+\infty$ whereas the antikink $K^{(-1,0)}_{\rm static}(\overline{x})$ reverses the previous path, see Figure 2. Notice that $K^{(-q,0)}_{\rm static}(\overline{x}) = \pi_x K^{(q,0)}_{\rm static}(\overline{x})$. The energy density (\ref{energydensity}) for the $K^{(q,0)}_{\rm static}(\overline{x})$-solutions is given by
\[
\epsilon [K^{(q,0)}_{\rm static}(\overline{x})] = \tau^2 \, \beta^2 \, {\rm sech}^4 (\tau \beta \, \overline{x}) \hspace{0.4cm},
\]
which corresponds to a localized energy density lump, see Figure 2. This implies that these topological defects can be interpreted as a first type of basic extended particles in the physical system. The total energy (\ref{totalenergy}) carried by these Type I solutions is
\begin{equation}
E[K^{(q,0)}_{\rm static}(\overline{x})] = \frac{4}{3} \, \tau \beta
\label{tk1totalenergy}
\end{equation}
The $K^{(1,0)}_{\rm static}(\overline{x})$-profile together with its energy density and its orbit have been displayed in Figure 2.

\begin{figure}[h]
\centerline{\includegraphics[height=3.cm]{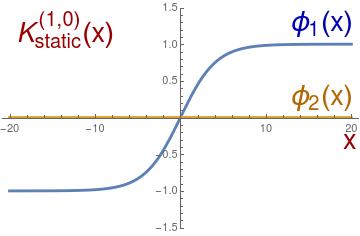} \hspace{0.3cm} \includegraphics[height=3.cm]{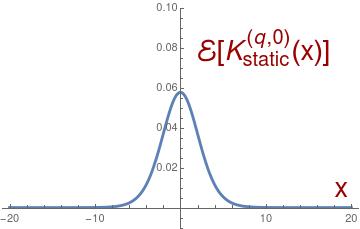} \hspace{0.3cm} \includegraphics[height=3.cm]{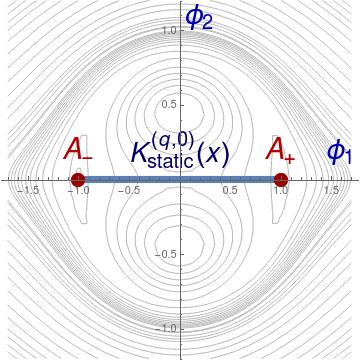}}
\caption{\small Graphics of the profile (left), energy density (middle) and orbit (right) of the $K^{(1,0)}_{\rm static}(\overline{x})$-kink with parameter values $\tau=1.2$ and $\beta=0.2$. A contour plot for the potential density $U(\phi_1,\phi_2)$ is used in the last figure. }
\end{figure}

The signs of the eigenvalues of the second order small $K^{(q,0)}_{\rm static}(\overline{x})$-kink fluctuation operator
\begin{equation}
{\cal H}[K^{(q,0)}_{\rm stat}(\overline{x})] =\left( \begin{array}{cc} -\frac{d^2}{dx^2} +4 \tau^2 \beta^2 - 6 \tau^2 \beta^2 \, {\rm sech}^2 (\tau \beta x) & 0 \\ 0 & -\frac{d^2}{dx^2} +\beta^2 \overline{\tau}^2 -2\tau^2 \beta^2 \,{\rm sech}^2( \tau \beta x) + {\rm sech}^4 (\tau \beta x)\end{array} \right) \label{tk1hess}
\end{equation}
determine the linear stability of this type of topological defects. In general, the analytical resolution of a spectral problem ${\cal H}[\Phi(x)] \psi_n = \omega_n^2 \psi_n$ associated with a matrix operator of the form
\begin{equation}
{\cal H}[\Phi(x)] = \left( \begin{array}{cc} -\frac{d^2}{dx^2} + V_{11}(x) & V_{12}(x) \\ V_{21}(x) & -\frac{d^2}{dx^2} +V_{22}(x) \end{array} \right) \label{genhess}
\end{equation}
is an unapproachable problem. In our framework, the potential well components $V_{ij}(x)$ for the second order small fluctuation operator ${\cal H}[\Phi(x)]$ are given by
\[
V_{ij}(x)=\frac{\partial^2 U}{\partial \phi_i \partial \phi_j}[\Phi(x)] \hspace{0.4cm}.
\]
For the one-component kinks the longitudinal and orthogonal $K^{(q,0)}_{\rm static}(\overline{x})$-fluctuations remain uncoupled
since $V_{12}(x)=V_{21}(x)=0$, see (\ref{tk1hess}). In Figure 3 (left), the potential wells $V_{ij}(x)$ of the ${\cal H}[K^{(q,0)}_{\rm static}(\overline{x})]$-operator for the parameter values $\tau=2.0$ and $\beta=0.4$ are displayed. Under the variable change $z= \tau \beta x$, the longitudinal eigenmodes $\psi_n^{\|}$ are described by the Schr\"odinger equation with a P\"oschl-Teller potential
\[
\Big[ -\frac{d^2}{dz^2} + 4 - 6 \, {\rm sech}^2 z \Big] \psi_n^{\|} = \frac{\omega_n^2}{\tau^2 \beta^2} \psi_n^{\|} \hspace{0.3cm},
\]
which is a solvable problem. The discrete spectrum comprises the usual translational zero mode and a vibrational eigenmode with $(\omega_1^2)^{\|}=\sqrt{3}\,\tau^2\beta^2$. In addition, a continuous spectrum emerges on the threshold value $(\omega_c^2)^{\|} =4\,\tau^2\beta^2$.

On the other hand, the orthogonal fluctuations $\psi_n^{\perp}$ are ruled by the spectral problem
\[
\Big[ -\frac{d^2}{dz^2} + \frac{\overline{\tau}^2}{\tau^2} - 2 \, {\rm sech}^2 z + \frac{1}{\tau^2 \beta^2} \, {\rm sech}^4 z \Big] \psi_n = \frac{\omega_n^2}{\tau^2 \beta^2} \psi_n \hspace{0.3cm},
\]
whose eigenvalues are not analytically known. Numerical analysis is employed in this case. The dependence on the coupling constant $\beta$ of the spectrum of the ${\cal H}[K^{(q,0)}_{\rm static}(\overline{x})]$-operator with $\tau=2.0$ is illustrated in Figure 3 (right). In general, a continuous spectrum associated with the orthogonal fluctuations arises on the value $(\omega_c^2)^\perp = \beta^2 \overline{\tau}^2$. For large enough values of $\beta$ a discrete eigenvalue $(\omega_1^2)^\perp$ emerges from the continuous spectrum, see Figure 3.

\begin{figure}[h]
\centerline{\includegraphics[height=3.cm]{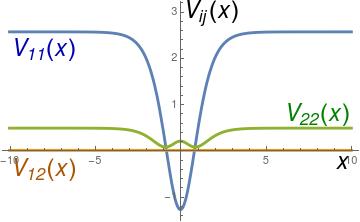}\hspace{1.0cm} \includegraphics[height=3.cm]{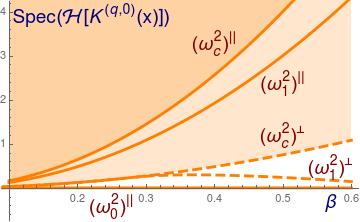}}
\caption{\small Graphics of the potential well components $V_{ij}(x)$ of the matrix operator (\ref{tk1hess}) for $\tau=2.0$ and $\beta=0.4$ (left) and dependence on the coupling constant $\beta$ of the ${\cal H}[K^{(q,0)}(\overline{x})]$-spectrum for the fixed value $\tau=2.0$ (right).}
\end{figure}

\noindent Under the assumptions (\ref{condi1}) and (\ref{condi2}), the second order small kink fluctuation operator ${\cal H}[K^{(q,0)}_{\rm static}(\overline{x})]$ comprises only non-negative eigenvalues. This implies the stability of the $K^{(q,0)}_{\rm static}(\overline{x})$-kinks.

\vspace{0.2cm}

\noindent (II) \textsc{Two-component topological kinks:} A second type of topological kinks describes the elliptic orbit 
\begin{equation}
\phi_1^2 + \tau^2 \phi_2^2 -1=0 \label{ellipse}
\end{equation}
in the internal plane $\phi_1-\phi_2$. Substituting this condition into the field equations (\ref{kleinequations}) leads to the topological kinks
\begin{equation}
{K}^{(q,\lambda)}_{\rm static}(\overline{x}) = \Big( q\, \tanh (\overline{\tau} \beta x)  , \frac{\lambda}{\tau} \, {\rm sech}\, (\overline{\tau}  \beta x) \Big) \label{tk2}
\end{equation}
where $q,\lambda=\pm 1$. Therefore, the expression (\ref{tk2}) defines four single solutions which join the vacua $A_\pm$. The magnitude $q$ is the topological charge, which distinguishes between kinks and antikinks and $\lambda$ determines whether the second kink component is positive or negative. Notice that ${K}^{(-q,\lambda)}_{\rm static}(\overline{x})= \pi_x {K}^{(q,\lambda)}_{\rm static}(\overline{x})$ and ${K}^{(q,-\lambda)}_{\rm static}(\overline{x})= \pi_2 {K}^{(q,\lambda)}_{\rm static}(\overline{x})$. In particular, the ${K}_{\rm static}^{(q,1)}(\overline{x})$-kinks live on the upper half-plane whereas the ${K}^{(q,-1)}_{\rm static}(\overline{x})$-kinks are confined to the lower half-plane, see Figure 4 (right). The components of the particular ${K}^{(1,1)}_{\rm static}(\overline{x})$-kink have been plotted in Figure 4 (left).

The energy density of these Type II solutions
\[
\epsilon[{K}^{(q,\lambda)}_{\rm static}(\overline{x})]=\frac{\overline{\tau}^2 \beta^2}{\tau^2} \, {\rm sech}^2(\overline{\tau} \, \beta x) \Big[ 1+ \overline{\tau}^2 \,{\rm sech}^2( \overline{\tau} \, \beta x) \Big]
\]
is localized around one point, see Figure 4 (middle). This means that there exists a second type of extended particles in the physical system. Indeed, it can be checked that the $K^{(q,0)}_{\rm static}(\overline{x})$-particles are more condensed than the ${K}^{(q,\lambda)}_{\rm static}(\overline{x})$-particles.

\begin{figure}[h]
\centerline{\includegraphics[height=3.cm]{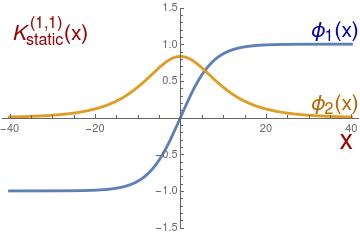} \hspace{0.3cm} \includegraphics[height=3.cm]{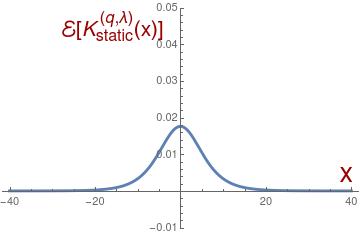} \hspace{0.3cm} \includegraphics[height=3.cm]{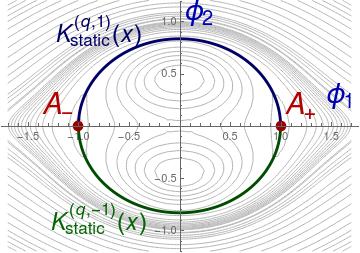}}
\caption{\small Graphics of the profile (left), energy density (middle) and orbit (right) of the ${K}^{(q,\lambda)}_{\rm static}(\overline{x})$-kinks with parameter values $\tau=1.2$ and $\beta=0.2$. A contour plot for the potential function $U(\phi_1,\phi_2)$ is used in last figure.}
\end{figure}

\noindent The total energy of the solutions (\ref{tk2}) is
\begin{equation}
E[{K}_{\rm static}^{(q,\lambda)}(\overline{x}) ] = \frac{2 \beta \overline{\tau}(1+2\tau^2)}{3 \tau^2} \hspace{0.3cm},
\label{tk2totalenergy}
\end{equation}
By comparing the expressions (\ref{tk1totalenergy}) and (\ref{tk2totalenergy}), it can be concluded that Type II extended particles are less energetic than Type I, that is,
\[
E[{K}_{\rm static}^{(q,\lambda)}(\overline{x}) ] < E[K^{(q,0)}_{\rm static}(\overline{x}) ] \hspace{0.3cm} , \hspace{0.3cm} \mbox{for } \lambda=\pm 1.
\]
The study of the linear stability in this case becomes a difficult task because the longitudinal and orthogonal ${K}^{(q,\lambda)}_{\rm static}(\overline{x})$-kink fluctuations are coupled by the operator (\ref{genhess}). Note that the potential well components $V_{ij}(x)$ are now given by the expressions
\begin{eqnarray}
V_{11}(x)&=& {\textstyle 4 \, \tau^2\beta^2 -\frac{2}{\tau^2} \, (-2-\tau^2\beta^2+3 \tau^4\beta^2) \,{\rm sech}^2 (\overline{\tau} \beta x) - \frac{4}{\tau^2}\, {\rm sech}^4 (\overline{\tau} \beta x) } \hspace{0.3cm}, \nonumber \\
V_{12}(x)&=& {\textstyle\frac{4}{\tau} \,{\rm sech}(\overline{\tau}\beta x) \, \tanh(\overline{\tau} \beta x) \left[\tau^2\beta^2 + \,{\rm sech}^2 (\overline{\tau} \beta x)\right]} \hspace{0.5cm}, \label{tk2hess} \\
V_{22}(x)&=& {\textstyle\beta^2 \overline{\tau}^2 -2(\tau^2-3) \beta^2 \, {\rm sech}^2 (\overline{\tau} \beta x) + 4 \, {\rm sech}^4(\overline{\tau} \beta x)} \hspace{0.3cm}. \nonumber
\end{eqnarray}
The behavior of these functions $V_{ij}(x)$ is shown in Figure 5 (left) for the parameter values $\tau=1.2$ and $\beta=0.2$. In Figure 5 (right) the dependence on the parameter $\beta$ of the spectrum of the operator ${\cal H}[{K}^{(q,\lambda)}_{\rm static}(\overline{x})]$ with $\tau=1.2$ (extracted by means of numerical analysis) is displayed. In general, the spectrum of the ${K}^{(q,\lambda)}_{\rm static}(\overline{x})$-fluctuation operator comprises a translational zero mode and two continuous spectra which emerge on the threshold values $4\tau^2 \beta^2$ and $\overline{\tau}^2 \beta^2$. No other discrete eigenvalue has been numerically identified in the regime determined by the conditions (\ref{condi1}) and (\ref{condi2}). Therefore, the lack of negative eigenvalues guarantees that the ${K}^{(q,\lambda)}_{\rm static}(\overline{x})$ kinks are stable solutions.

\begin{figure}[h]
\centerline{\includegraphics[height=3.cm]{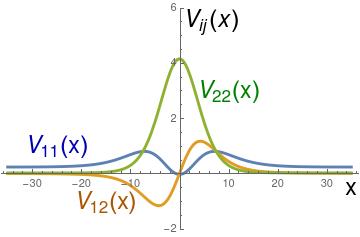}\hspace{1.0cm} \includegraphics[height=3.cm]{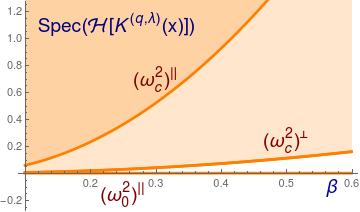}}
\caption{\small Graphics of the potential well components (\ref{tk2hess}) for the parameter values $\tau=1.2$ and $\beta=0.2$ (left) and dependence on the coupling constant $\beta$ of the ${\cal H}[{K}^{(q,\lambda)}_{\rm static}(\overline{x})]$-spectrum for the fixed value $\tau=1.2$ (right).}
\end{figure}

\noindent In sum, the two-component scalar field theory model introduced in this Section involves two different types of basic extended particles:
\begin{enumerate}
	\item \textit{Type I particles}. These extended particles are described by one-component kinks $K_{\rm static}^{(q,0)} (x)$, which are specified by the analytical expression (\ref{tk1}). Two  different possible values of the topological charge $q=\pm 1$ are carried by these particles. In this sense, Type I extended particles with $q=-1$ can be thought of as Type I extended antiparticles of those with positive topological charge.
	
	\item \textit{Type II particles}. The two-component topological kinks $K_{\rm static}^{(q,\lambda)}(x)$ given by (\ref{tk2}) for $\lambda=\pm 1$ describe this class of particles. These energy lumps are characterized by the value of the charge pair $(q,\lambda)$. Again, the topological charge $q$ distinguishes between particles and antiparticles. Analytically, the value of $\lambda$ determines if the second component of the kink profile is positive or negative. From the physical perspective this number can be interpreted as a new property of the extended particles. As we shall see later, Type II particles with different charge $\lambda$ interact very differently than those with the same $\lambda$ when they collide each other.
\end{enumerate}

For the sake of simplicity, two different index symbols will be employed from now on: $\Lambda$ shall denote an index whose possible values are $\Lambda=0,\pm 1$ whereas $\lambda$ is restricted to the values $\pm 1$, that is, $\lambda=\pm 1$. In this way,
\[
K_{\rm static}^{(q,\Lambda)}(x) \hspace{1cm} \mbox{where } \Lambda=0,\pm 1, \hspace{0.4cm} q=\pm 1,
\]
represents the set of all the previous kink solutions whereas
\[
K_{\rm static}^{(q,\lambda)}(x) \hspace{1cm} \mbox{where } \lambda=\pm 1, \hspace{0.5cm} q=\pm 1,
\]
refers only to the two-component topological kinks.

\section{Kink scattering}

\noindent In this Section, the study of the scattering between the extended particles identified in the previous Section is addressed. Therefore, the kink dynamics derived from the evolution equations (\ref{kleinequations}) must be analyzed for colliding kink configurations. Static kinks introduced in Section 2 can be transformed into constant velocity traveling kinks
\begin{equation}
K^{(q,\Lambda)}(x,t;v_0) = K_{\rm static}^{(q,\Lambda)} \Big( \frac{\overline{x}-v_0 t}{\sqrt{1-v_0^2}} \Big) \hspace{0.4cm},\hspace{0.4cm} q=\pm 1, \hspace{0.2cm} \Lambda=0,\pm 1 \hspace{0.5cm},
\label{travelingkink}
\end{equation}
by using the Lorentz invariance of the action (\ref{action}). The solutions (\ref{travelingkink}) of the equations (\ref{kleinequations}) describe traveling extended particles. By using these expressions, the initial configurations for our scattering problems will be constructed by concatenating  two well-separated kinks, which approach each other with velocity $v_0$, that is,
\begin{equation}
K^{(q,\Lambda_1)}(x+x_0,t;v_0) \cup K^{(-q,\Lambda_2)}(x-x_0,t;-v_0) \hspace{0.2cm}, \hspace{0.8cm} \Lambda_1,\Lambda_2=0,\pm 1, \hspace{0.2cm} \label{generalconca}
\end{equation}
where $x_0$ is large enough to guaranty the smoothness of the initial configuration. Notice that these well-separated kinks must carry opposite topological charges. Therefore, the initial multi-kink configuration carries zero topological charge. It asymptotically begins and ends at the same vacuum point. Taking into account this fact and the system symmetries, the catalog of possible two-kink scattering events in this model is given as follows:

\begin{itemize}
\item[(a)] $K^{(q,0)}(x,v_0)- K^{(-q,0)}(x,-v_0)$ \textit{scattering processes}. This class of events involves the collision between a Type I particle and its antiparticle, or in other words, the scattering between a one-component kink and its antikink. The second component of the solutions is always zero, so the second equation in (\ref{kleinequations}) is automatically satisfied. The problem is reduced, therefore, to the kink scattering in the one-component $\phi^4$ model. In this case, it is well known that if the initial velocity $v_0$ is greater than the critical speed $v_c\approx 0.2598$ the single solutions reflect each other but if $v_0<v_c$ then they collide a second time. In this case the usual result is the formation of a bion except for some velocity windows (exhibiting a fractal structure) where the kinks escape after a finite number of collisions. Therefore, the collision between a Type I particle and its antiparticle leads to the annihilation or reflection of the same particles. This class of scattering processes will not be dealt with in this work (see, for instance, the seminal work \cite{Campbell1983} for details) because we are interested in exploring new phenomena in multi-component kink collisions.

\item[(b)] $K^{(q,\lambda)}(x,v_0)- K^{(-q,\lambda)}(x,-v_0)$ \textit{scattering processes} with $\lambda=\pm 1$. Collisions between Type II particle-antiparticle pairs are encompassed in this category. A two-component kink and its own antikink, both with the same charge $\lambda$, are pushed with impact velocity $v_0$. The initial multi-kink profile is plotted in Figure 6 (left). It consists of two pieces: a $K^{(q,\lambda)}(x,v_0)$-kink (represented by a solid line) for $x<0$ and an antikink $K^{(-q,\lambda)}(x,-v_0)$ (represented by a dashed line) for $x>0$. This initial configuration starts at the vacuum $A_{-q}$, follows a semi-elliptic trajectory in the internal half-plane $(-1)^{(\lambda-1)/2} \phi_2>0$, approaches to the point $A_q$ and finally returns to the initial vacuum $A_{-q}$ by reversing the previous orbit, see Figure 6 (right). This multi-kink configuration is not affected by any potential barrier and presumably its initial evolution will tend to the vacuum configuration $A_{-q}$, at least for small collision velocities. As we shall see in Section 3.3, if the initial velocity $v_0$ is large enough the creation of a particle/antiparticle pair is also possible.

\begin{figure}[h]
	\centerline{\includegraphics[height=3.cm]{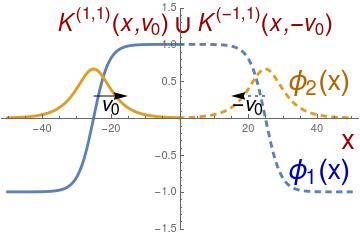}\hspace{1.0cm} \includegraphics[height=3.cm]{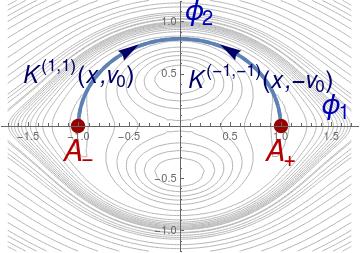}}
	\caption{\small Initial configuration for the $K^{(q,\lambda)}(x,v_0)- K^{(-q,\lambda)}(x,-v_0)$ scattering processes:  Multi-kink profile for the first and second component of the scalar field  $\phi(x)$ (left) and initial multikink orbit in the internal plane (right). A contour plot for the potential density $U(\phi_1,\phi_2)$ is used in the last figure.}
\end{figure}

\item[(c)] $K^{(q,\lambda)}(x,v_0)- K^{(-q,-\lambda)}(x,-v_0)$ \textit{scattering processes} with $\lambda=\pm 1$. This class of events also involves collisions between two Type II extended particles, but now the two-component kinks in (\ref{generalconca}) carry different charge $\lambda$. In these circumstances, the $ K^{(-q,-\lambda)}(x)$-solution is not the antikink of the $K^{(q,\lambda)}(x)$-kink. The initial configuration starts and ends at the point $A_{-q}$ describing a complete elliptic orbit that passes closely through the point $A_{q}$, see Figure 7. Note that this arrangement encloses the origin of the internal plane, which means that there exists a potential barrier between the two well-separated topological defects. It will be shown in Section 3.2 that the kink dynamics in this case is completely different from that in the case (b) despite the fact that both of them involve collisions between Type II particles.

\begin{figure}[h]
	\centerline{\includegraphics[height=3.cm]{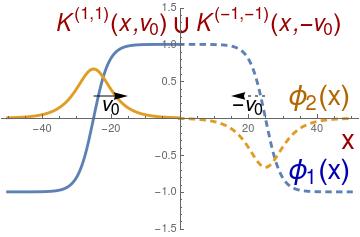}\hspace{1.0cm} \includegraphics[height=3.cm]{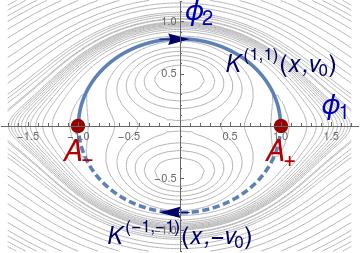}}
	\caption{\small Initial configuration for the $K^{(q,\lambda)}(x,v_0)- K^{(-q,-\lambda)}(x,-v_0)$ scattering processes: Multi-kink profile for the first and second component of the scalar field  $\phi(x)$ (left) and initial multikink orbit in the internal plane (right). A contour plot for the potential density $U(\phi_1,\phi_2)$ is used in the last figure.}
\end{figure}

\item[(d)] $K^{(q,\lambda)}(x,v_0)- K^{(-q,0)}(x,v_0)$ \textit{scattering processes}  with $\lambda=\pm 1$. Finally, the last class of two-kink scattering events which will be addressed in this work concerns collisions between Type I and II particles. Given that the colliding particles have different nature it is expected these scattering processes to evolve asymmetrically. In particular, a two-component $K^{(q,\lambda)}(x)$-kink and an one-component $K^{(-q,0)}(x)$-solution are obliged to collide each other, see Figure 8 (left). By convention, the two-component kink $K^{(q,\lambda)}(x)$ (represented by a solid line) is initially placed to the left of the one-component kink $K_{\rm static}^{(-q,0)}(x)$ (represented by a dashed line) in the spatial axis $x$. The use of system symmetries allows us to analyze other initial arrangements on the basis of the results obtained in this case. The orbit of this multikink configuration describes a semi-ellipse which starts at the vacuum $A_{-q}$ and arrives to the point $A_{q}$, and later returns to the vacuum $A_{-q}$ following a straight line placed on the axis $\phi_1$, see Figure 8 (right).  The last piece of the multikink trajectory crosses the origin of the internal plane. Again, a potential barrier between the two concatenated kinks arises, although now it is less strong than in the case (c). Besides, the larger the value of $\tau$ is the weaker this barrier is, see Figure 1. We shall discuss these processes in Section 3.1.

\begin{figure}[h]
	\centerline{\includegraphics[height=3.cm]{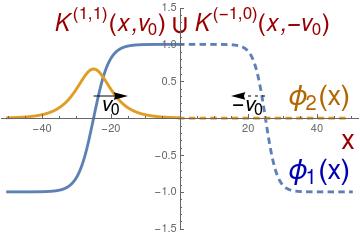}\hspace{1.0cm} \includegraphics[height=3.cm]{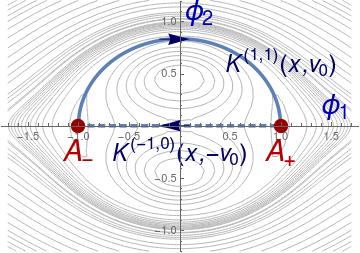}}
		\caption{\small Initial configuration for the $K^{(q,\lambda)}(x,v_0)- K^{(-q,0)}(x,-v_0)$ scattering processes: Multi-kink profile for the first and second component of the scalar field  $\phi(x)$ (left) and initial multikink orbit in the internal plane (right). A contour plot for the potential density $U(\phi_1,\phi_2)$ is used in the last figure.}
\end{figure}
\end{itemize}

The study of the previous kink scattering processes demands the analysis of the evolution of the initial multi-kink configuration derived from the equations (\ref{kleinequations}). The resulting configuration after the kink collision is characterized by the resulting particles emerging after the impact together with the value of its final velocities.
The final output of these events critically depends on several factors: (1) the type of scattering processes,  (2) the initial collision velocity $v_0$ of the colliding kinks and (3) the particular values of the coupling constants $\tau$ and $\beta$ of the model. However, the topological charge of all the final configurations must be zero because this magnitude is a system invariant. This fact allows us to establish the main scattering channels for the previous events:

\begin{itemize}
	\item[(1)] \textit{Mutual annihilation}: $K^{(q,\Lambda_1)}(v_0) \cup K^{(-q,\Lambda_2)}(-v_0) \rightarrow \nu$ .
	
	In this situation the kink $K^{(q,\Lambda_1)}(x)$ and antikink $K^{(-q,\Lambda_2)}(x)$ approach each other, collide and finally annihilate each other giving rise to a radiation vestige in the space. The final configuration consists of a packet of fluctuations around one of the vacua. This ultimate process can be reached as the result of the evolution of a bion, where kink and antikink repeatedly bounce and radiate energy in every collision.
	
	\item[(2)] \textit{Emission of a kink-antikink pair}: $K^{(q,\Lambda_1)}(v_0) \cup K^{(-q,\Lambda_2)}(-v_0)  \rightarrow K^{(q,\Lambda_3)}(-v_f)\cup K^{(-q,\Lambda_4)}(\overline{v}_f)$.
	
	In this type of events, a kink-antikink pair emerges after the $K^{(q,\Lambda_1)}(v_0) -K^{(-q,\Lambda_2)}(-v_0)$-collision. The final charges $\Lambda_3$ and $\Lambda_4$ of the emerging particles are not only fixed by the charges $\Lambda_1$ and $\Lambda_2$ of the colliding particles but they also depend on the velocity $v_0$ of these initial lumps. The same scattering event can produce distinct particle/antiparticle pairs for different collision velocities. It is assumed that the $K^{(q,\Lambda_3)}(-v_f)$-kink will travel to the left with final velocity $v_f$ whereas the $K^{(-q,\Lambda_4)}(\overline{v}_f)$-antikink will travel to the right with final velocity $\overline{v}_f$. Radiation can also be emitted in these processes. Taking into account that $\Lambda_3,\Lambda_4=0,\pm 1$, nine different final scenarios are possible, which we classify in the following points:
	\begin{enumerate}
		\item[(2a)] If $\Lambda_3=\Lambda_4=0$, a one-component kink-antikink pair emerges after the collision of the initial lumps. As a result, two Type I extended particles are created and move away in the spatial axis.
		
		\item[(2b)] If $\Lambda_3,\Lambda_4=\pm 1$, the original kinks collide and transform into the most energetically favorable  configuration formed by a pair of Type II extended particles. If $\Lambda_3=\Lambda_4$ then the final configuration involves a two-component kink-antikink pair but if $\Lambda_3\neq \Lambda_4$ then the kink and the antikink carry different $\lambda$-charge, so that its orbit describes the complete ellipse $\phi_1^2+\tau^2\phi_2^2=1$.
		
		\item[(2c)] Finally, if $|\Lambda_3|=1$ and $\Lambda_4=0$ or $\Lambda_3=0$ and $|\Lambda_4|=1$ then an asymmetric situation takes place where a Type I extended particle moves away from a Type II one. In the first case the Type II particle is placed to the left of the Type I particle whereas the order is reversed in the second case.
	\end{enumerate}
\end{itemize}

The number of possible final configurations (previously discussed) shows the complexity of the kink scattering in this model. The initial velocity $v_0$ plays an essential role in the kink scattering processes. Indeed, the nature of the resulting topological defects as well as its final velocities $v_f$ can be quite different even for close initial velocities $v_0$. We shall begin by studying the asymmetric kink scattering processes (d) in Subsection 3.1. Obviously, due to the non-linearity of the evolution equations (\ref{kleinequations}), any attempt to analytically solve the problem is unsuccessful in this case. Numerical analysis will be employed to study the evolution of the initial multi-kink configuration $K^{(q,\Lambda_1)}(v_0) \cup K^{(-q,\Lambda_2)}(-v_0)$. The numerical procedure used in this paper follows the algorithm described in \cite{Kassam2005} by Kassam and Trefethen. This scheme is spectral in space and fourth order in time and was designed to solve the numerical instabilities of the exponential time-differencing Runge-Kutta method introduced in \cite{Cox2002}. As a complement to this numerical method, an energy conservative second-order finite difference algorithm \cite{Alonso2017} implemented with Mur boundary conditions \cite{Mur1981} has also been employed. The effect of radiation in the simulation is controlled by this algorithm because the linear plane waves are absorbed at the boundaries. The two previous numerical schemes provide similar results. The symmetric kink scattering events described in the points (b) and (c) will be dealt with in Subsection 3.2 and 3.3.

\subsection{${K}^{(q,\lambda)}(x)$-$K^{(-q,0)}(x)$ scattering processes }

In this subsection the asymmetric scattering processes given by collisions between a two-component kink $K^{(q,\lambda)}(x,t;v_0)$ and an one-component antikink $K^{(-q,0)}(x,t;-v_0)$ are numerically investigated. The initial configuration is represented by the concatenation
\begin{equation}
K^{(q,\lambda)}(x-x_0,t;v_0) \cup K^{(-q,0)}(x+x_0,t;-v_0) \label{conca12} \hspace{0.3cm} ,
\end{equation}
where the single kinks approach each other with speed $v_0$, see Figure 8.

The kink scattering results for this type of events have been summarized in Figure 9 for three different values of the model parameters. In these graphics the final velocities $v_f$ of the left-traveling kink and $\overline{v}_f$ of the right-traveling antikink are plotted as a function of the initial collision velocity $v_0$. A zero final velocity means that an annihilation process has taken place and only radiation remains. On the other hand, the identity of the scattered kinks has been represented by a color stripe pattern. The color code (distinguishing the ten possible final scenarios) has been included in Figure 9.

\begin{figure}[h]
	\centerline{\begin{tabular}{c}
			\includegraphics[height=2.cm]{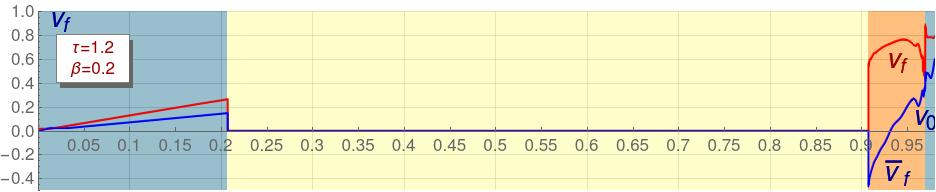}\\
			\includegraphics[height=2.cm]{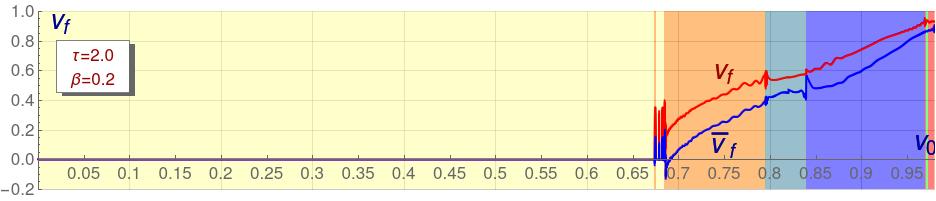}\\
			\includegraphics[height=2.cm]{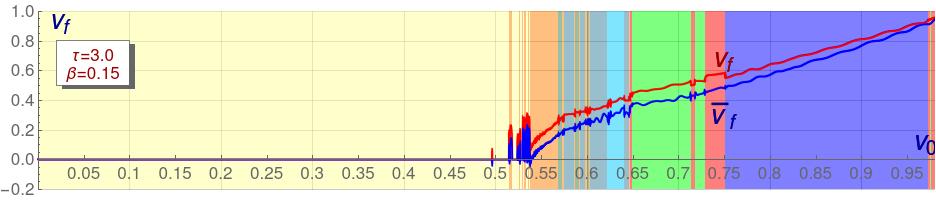}  \end{tabular} \\ \hspace{0.3cm}
		\begin{tabular}{c} \includegraphics[height=5.5cm]{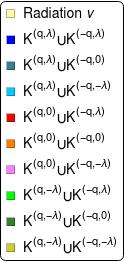}\end{tabular}}
	\caption{\small Graphical representation of the final velocities $v_f$ and $\overline{v}_f$ of the kink and the antikink as a function of the initial velocity $v_0$ for the ${K}^{(q,\lambda)}(x)$-$K^{(-q,0)}(x)$ scattering processes in the cases $\tau=1.2$, $\beta=0.2$ (left, top), $\tau=2.0$, $\beta=0.2$ (left, middle) and $\tau=3.0$, $\beta=0.15$ (left, bottom). A color code distinguishing the final scenarios has been included (right).}
\end{figure}

The results illustrated in Figure 9 are described in the following points:
\begin{enumerate}
	\item For the coupling constants $\tau=1.2$ and $\beta=0.2$, the potential barrier between the two single colliding kinks is strong enough to establish an elastic regime for low initial velocities $v_0$, see Figure 9 (top). This type of scattering processes can be represented as
	\begin{equation}
	K^{(q,\lambda)}(v_0) \cup K^{(-q,0)}(-v_0) \rightarrow K^{(q,\lambda)}(-v_f) \cup K^{(-q,0)}(\overline{v}_f) \hspace{0.5cm} . \label{processd1}
	\end{equation}
	Here, the particles approach each other but when they are close enough the lumps repel each other and move away without radiation emission. If the initial velocity belongs to the interval $v_0\in [0.21,0.9076]$ the potential barrier between the original kinks is overcome and an annihilation regime arises. The Type I and Type II extended particles collide and mutually destroy. This situation is characterized by the process
	\begin{equation}
	K^{(q,\lambda)}(v_0) \cup K^{(-q,0)}(-v_0) \rightarrow \nu \hspace{0.5cm} .  \label{processd2}
	\end{equation}
	For larger values of the collision velocities, the potential barrier can be surpassed twice and a kink-antikink pair emerges again, see Figure 9 (top). For the initial velocity window $v_0\in [0.9076,0.9696]$, two Type I extended particles (described by one-component kinks) are created after the impact, which involves the event
	\begin{equation}
	K^{(q,\lambda)}(v_0) \cup K^{(-q,0)}(-v_0) \rightarrow K^{(q,0)}(-v_f) \cup K^{(-q,0)}(\overline{v}_f) + \nu \hspace{0.5cm} .  \label{processd3}
	\end{equation}
	In this regime a large amount of kinetic energy is converted into radiation. Negative values of the final velocity $\overline{v}_f$ can be found for some values of the initial velocity. This means that the two extended particles travel toward the left for these cases. For greater values of the initial velocity an inelastic kink reflection described by the process (\ref{processd1}) occurs. In this case a Type I and Type II extended particles emerge with the same charges than the colliding ones although now a large amount of radiation is emitted. 	
	
	\item For the case $\tau=2.0$ and $\beta=0.2$, the potential barrier between kink and antikink is so weak that the elastic regime is almost indiscernible. The kink collision provokes the annihilation (\ref{processd2}) of the two extended particles for values of the initial velocity less than $v_0\approx 0.6857$, see Figure 9 (middle). Again, this regime is followed by collision velocity windows where two extended particles emerge after the kink impact. The first such windows leads to the scattering processes (\ref{processd2}) where two Type I extended particles are created. The transition between the annihilation regime and this one is diffuse due to the presence of resonance windows. For these initial velocities the extended particles remain bound but after a finite number of collisions they are able to escape because they capture kinetic energy from the vibrational modes due to the resonant energy transfer mechanism. A three-bounce scattering event is illustrated in Figure 10. A Type I and a Type II particles approach, collide and emerge as two Type I particles; these lumps collide twice more and eventually escape and move away.
	
	\begin{figure}[h]
		\centerline{\includegraphics[height=3.cm]{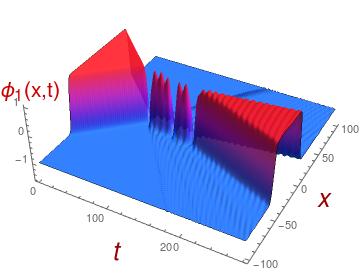} \hspace{1.5cm} \includegraphics[height=3.cm]{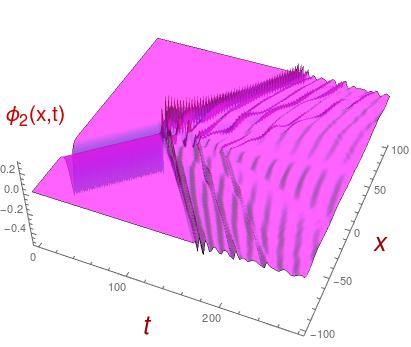} }
		\caption{\small Evolution of the first and second scalar field components for a ${K}^{(q,\lambda)}(x)$-$K^{(-q,0)}(x)$ scattering process with impact velocity $v_0=0.6785$ and model parameters $\tau=2.0$ and $\beta=0.2$.}
	\end{figure}

	The next regime is defined approximately in the interval $v_0\in [0.795,0.840]$ and is characterized by an inelastic kink reflection (\ref{processd1}) where the original extended particles finally emerge and separate each other. A new class of scattering events appears for the interval $v_0\in [0.840,0.969]$. Now a Type II extended particle/antiparticle pair is created after the kink impact, giving rise to the event 
	\begin{equation}
	K^{(q,\lambda)}(v_0) \cup K^{(-q,0)}(-v_0) \rightarrow K^{(q,\lambda)}(-v_f) \cup K^{(-q,\lambda)}(\overline{v}_f)+ \nu \hspace{0.5cm} . \label{processd4}
	\end{equation}
	The transitions between these regimes can be distinguished by the presence of quasi-resonances where the final velocities of the scattered kinks dramatically drop or change its tendency, see Figure 9 (middle). For very large speeds other velocity bands with very small widths arise giving rise to other final kink configurations.
	
\item For $\tau=3.0$ and $\beta=0.15$ the pattern is more complex. As before, an annihilation regime is followed by an initial velocity interval where a Type I particle/antiparticle pair is created after the collision, see Figure 9 (bottom). These regimes are separated by resonance windows, where the kink and the antikink collide and bounce back a finite number of times before escaping. For greater values of $v_0$ a complex sequence of initial velocity bands arises where several scattering processes are alternated. For example, in the range $v_0\in [0.6235,0.6406]$ events of the form
	\begin{equation}
	K^{(q,\lambda)}(v_0) \cup K^{(-q,0)}(-v_0) \rightarrow K^{(q,\lambda)}(-v_f) \cup K^{(-q,-\lambda)}(\overline{v}_f) + \nu \label{processd5}
	\end{equation}
take place. This type of processes can be interpreted as follows: the original Type I and Type II extended particles collide and bounce back while as the same time the Type I particle decays to a Type II extended particle whose $\lambda$-charge is opposite to that of the colliding lump. For the range $v_0\in [0.65,0.713] \cup [0.718,0.729]$ the scattering processes
	\begin{equation}
	K^{(q,\lambda)}(v_0) \cup K^{(-q,0)}(-v_0) \rightarrow K^{(q,-\lambda)}(-v_f) \cup K^{(-q,\lambda)}(\overline{v}_f) + \nu \label{processd6}
	\end{equation}
occur. The interpretation in this case is more subtle: Type I and Type II extended particles collide and bounce back, the left-traveling Type II particle reverses its $\lambda$-charge and the right-traveling Type I particle decays to a Type II particle with the same $\lambda$-charge than the colliding one. This class of events has been graphically represented in Figure 11. 
    
\begin{figure}[h]
\centerline{\includegraphics[height=3.cm]{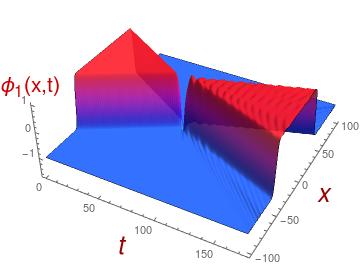} \hspace{1.5cm} \includegraphics[height=3.cm]{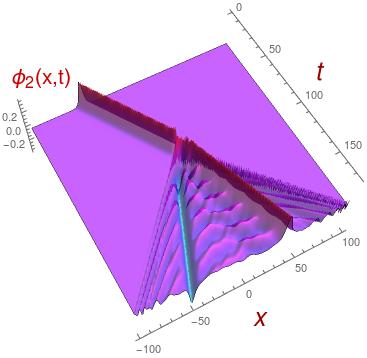} }
\caption{\small Evolution of the first and second scalar field components for a ${K}^{(q,\lambda)}(x)$-$K^{(-q,0)}(x)$ scattering process with impact velocity $v_0=0.68$ and model parameters $\tau=3.0$ and $\beta=0.15$.}
\end{figure}

Observe the evolution of the second scalar field in this process, see Figure 11. An initial configuration exhibiting only one peak is transformed into a two-peak configuration together with radiation. In Figure 12, the evolution of the multi-kink orbit is depicted for several values of the time. This is a very illustrative sequence of graphics, which shows the complexity of the kink collisions in two-component scalar field theory models.

\begin{figure}[h]
	\centerline{\includegraphics[height=1.6cm]{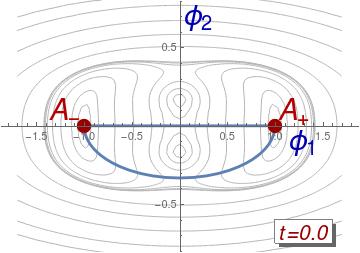}
		\includegraphics[height=1.6cm]{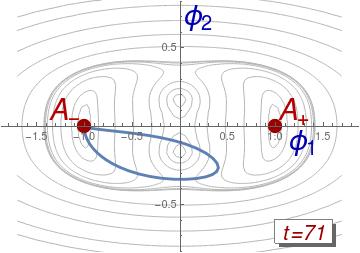}
		\includegraphics[height=1.6cm]{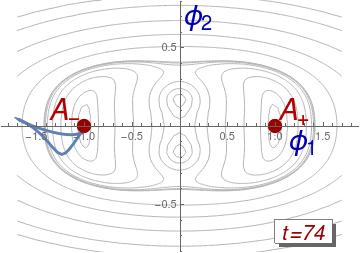}
		\includegraphics[height=1.6cm]{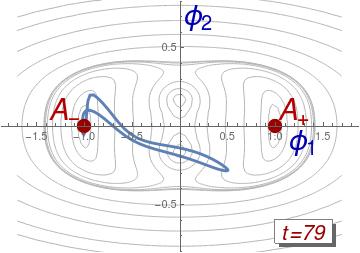}
		\includegraphics[height=1.6cm]{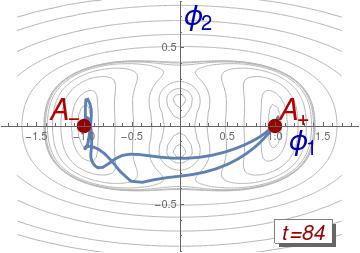}
		\includegraphics[height=1.6cm]{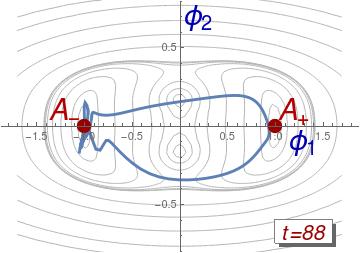}
		\includegraphics[height=1.6cm]{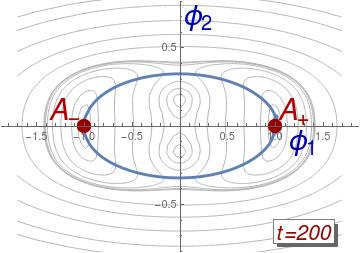}  }
	\caption{\small  Evolution of multi-kink orbit (\ref{conca12}) for a ${K}^{(q,\lambda)}(x)$-$K^{(-q,0)}(x)$ scattering process with impact velocity $v_0=0.68$ and model parameters $\tau=3.0$ and $\beta=0.15$.}
\end{figure}

Scattering processes which generate a Type II particle/antiparticle pair arise approximately in the interval $v_0\in [0.751,0.973]$, see Figure 9 (bottom). Other more narrow bands involve the creation of a Type I particle/antiparticle pair or the creation of different combinations of a Type I and a Type II traveling particles, see Figure 9 (bottom).
\end{enumerate}

\subsection{${K}^{(q,\lambda)}(\overline{x})$-${K}^{(-q,-\lambda)}(\overline{x})$ scattering processes }

In this subsection, scattering processes between a kink $K^{(q,\lambda)}(x,t;v_0)$ and an antikink $K^{(-q,-\lambda)}(x,t;-v_0)$ are numerically investigated. The initial configuration consists of two well separated boosted static kinks
\begin{equation}
K^{(q,\lambda)}(x-x_0,t;v_0) \cup K^{(-q,-\lambda)}(x+x_0,t;-v_0) \label{conca2} \hspace{0.3cm} ,
\end{equation}
which are pushed together with speed $v_0$. The orbit of the kink-antikink configuration (\ref{conca2}) describes an ellipse, which starts and ends at the vacuum point $A_{- q}$, see Figure 7. The first component of the initial configuration (\ref{conca2})
is symmetric with respect to the spatial reflection $\pi_x$ whereas the second component is antisymmetric. The final kink configuration must preserve these symmetries, which implies that the scattering processes (2c) involving the creation of a Type I and Type II extended particles are forbidden. Besides, the emerging lumps must carry opposite $\Lambda$-charges. For these symmetric events, only one final velocity $v_f$ is needed to determine the speed of the scattered kinks. In this case, the mass center can be chosen to be at the origin of the spatial axis $x$. The scattering data for this type of events are graphically summarized in Figure 13 for the cases $\tau=2.0$ and $\tau=3.0$ with $\beta=0.2$. The nature of the resulting topological defects as well as its separation velocity $v_f$ depend on the initial velocity $v_0$. A general description of the found pattern is given as follows:

\begin{figure}[h]
	\centerline{\begin{tabular}{c}
			\includegraphics[height=2.cm]{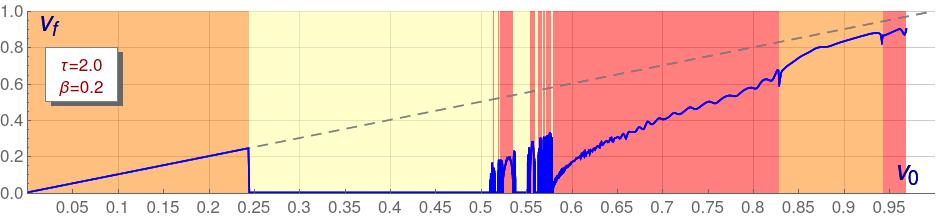}\\
			\includegraphics[height=2.cm]{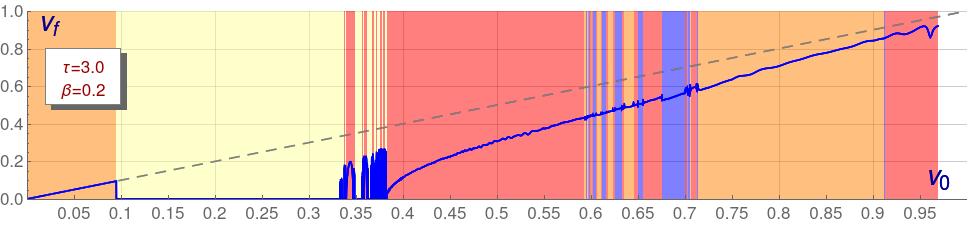} \end{tabular} \\ \hspace{0.3cm}
		\begin{tabular}{c} \includegraphics[height=2.cm]{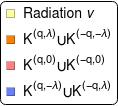}\end{tabular}}
	\caption{\small Graphical representation of the final velocity $v_f$ of the kink and the antikink as a function of the initial velocity $v_0$ for the ${K}^{(q,\lambda)} (\overline{x})$-${K}^{(-q, -\lambda)} (\overline{x})$ scattering processes in the cases $\tau=2.0$, $\beta=0.2$ (left, top) and $\tau=3.0$, $\beta=0.2$ (left, bottom). For the sake of comparison a dashed straight line has been added in these figures representing the final velocity for elastic collisions. A color code distinguishing the possible events has been attached (right). }
\end{figure}

\noindent -- (1) For small values of the initial velocity $v_0$ the scattering is approximately elastic. The kink and the antikink approach each other and when they are close enough repulsive forces arise between these energy lumps. The kink and the antikink repel each other and finally move away with approximately the same initial velocity $v_0$. The resulting topological defects after the collision coincide with the original ${K}^{(q,\lambda)}(\overline{x})$ and ${K}^{(-q,-\lambda)}(\overline{x})$ kinks. This type of processes can be symbolically represented as
\begin{equation}
K^{(q,\lambda)}(v_0) \cup K^{(-q,-\lambda)}(-v_0) \rightarrow K^{(q,\lambda)}(-v_0) \cup K^{(-q,-\lambda)}(v_0)   \label{pheno1}
\end{equation}

\vspace{0.2cm}

\noindent -- (2) If the initial velocity $v_0$ increases the repulsive forces between the $K^{(q,\lambda)}(x)$ and $K^{(-q,-\lambda)}(x)$ are overcome and the extended particles collide each other. Now, a bound state (bion)
is formed, however, this bion does not consist of the original $K^{(q,\lambda)}(x)$ and $K^{(-q,-\lambda)}(x)$ kinks (whose second components are non null) but of the kink $K^{(q,0)}(x)$ and its antikink $K^{(-q,0)}(x)$ (whose second components vanish). This pair of new lumps is forced to approach and bounce back over and over again while emitting radiation. Eventually, the sustained radiation emission can provoke the kink-antikink annihilation. This type of phenomena can be symbolized as
\begin{equation}
K^{(q,\lambda)}(v_0) \cup K^{(-q,-\lambda)}(-v_0) \rightarrow K^{(q,0)} \uplus K^{(-q,0)} + \nu \label{pheno2}
\end{equation}
where the symbol $\uplus$ stands for the formation of a bion. Figure 14 displays the evolution of the configuration (\ref{conca2}) for a collision velocity $v_0=0.4$ and model parameters $\tau=2.0$ and $\beta=0.2$. 

\begin{figure}[h]
\centerline{\includegraphics[height=3.cm]{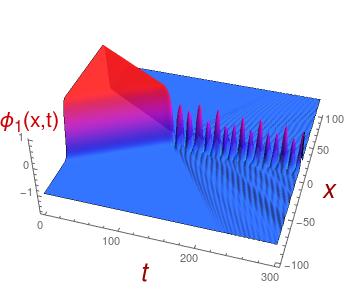} \hspace{1.5cm} \includegraphics[height=3.cm]{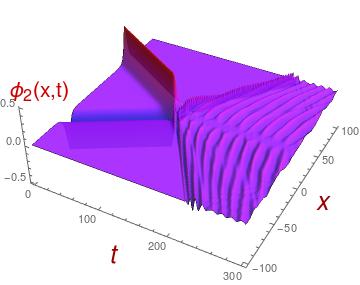} }
\caption{\small Evolution of the first and second scalar field components for a $K^{(q,\lambda)}(v_0)$-$K^{(-q,-\lambda)}(-v_0)$ scattering process with impact velocity $v_0=0.4$ and model parameters $\tau=2.0$ and $\beta=0.2$.}
\end{figure}

\vspace{0.2cm}

\noindent -- (3) For larger values of the initial velocity $v_0$, there exist events for which the $K^{(q,0)}(-v_f)$-kink and its antikink emerge after the collision between the $K^{(q,\lambda)}(v_0)$-kink and the  $K^{(-q,-\lambda)}(-v_0)$-antikink. This type of scattering processes, which is represented as
\begin{equation}
K^{(q,\lambda)}(v_0) \cup K^{(-q,-\lambda)}(-v_0) \rightarrow K^{(q,0)}(-v_f) \cup K^{(-q,0)}(v_f)  + \nu \label{pheno3}
\end{equation}
is illustrated in Figure 15. The bell-shape dependence of the second field component carried by the topological defects $K^{(q,\lambda)}(x)$ and $K^{(-q,-\lambda)}(x)$ is destroyed and converted into small fluctuation around the value $\phi_2=0$, see Figure 15. In addition, the profile of the first field component now becomes sharper. Excitation of internal vibrational eigenmodes together with radiation emission are also involved in these processes, which implies that $v_f<v_0$. 

\begin{figure}[h]
\centerline{\includegraphics[height=3.cm]{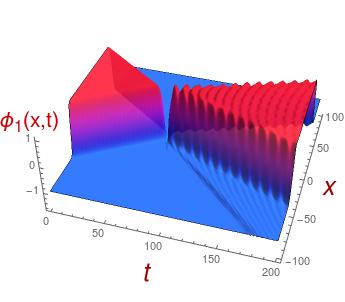} \hspace{1.5cm} \includegraphics[height=3.cm]{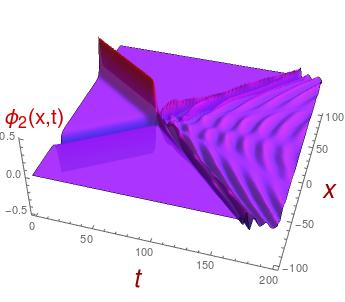} }
\caption{\small  Evolution of the first and second scalar field components for a $K^{(q,\lambda)}(v_0)$-$K^{(-q,-\lambda)}(-v_0)$ scattering process with impact velocity $v_0=0.7$ and model parameters $\tau=2.0$ and $\beta=0.2$.}
\end{figure}

In Figure 16, the evolution of the orbit in the process illustrated in Figure 15 has been plotted for different times. The original elliptic trajectory (\ref{conca2}) is deformed into a smaller closed curve, which later evolves to a path described very closely by a straight line joining the vacuum points $A_\pm$, which is traversed twice.

\begin{figure}[h]
\centerline{\includegraphics[height=1.6cm]{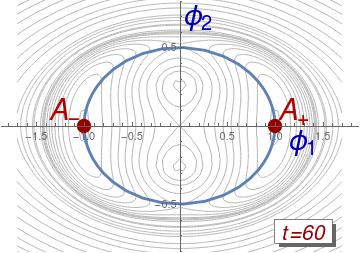}
\includegraphics[height=1.6cm]{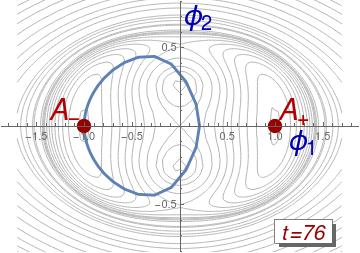}
\includegraphics[height=1.6cm]{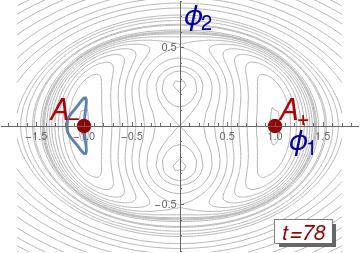}
\includegraphics[height=1.6cm]{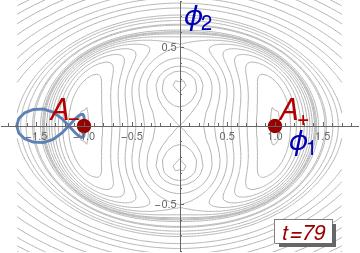}
\includegraphics[height=1.6cm]{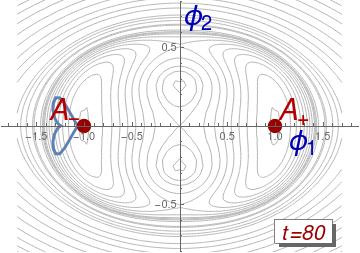}
\includegraphics[height=1.6cm]{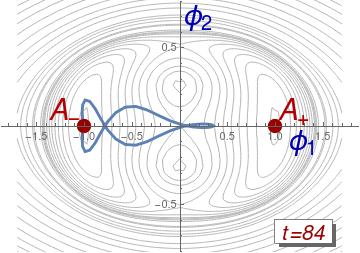}
\includegraphics[height=1.6cm]{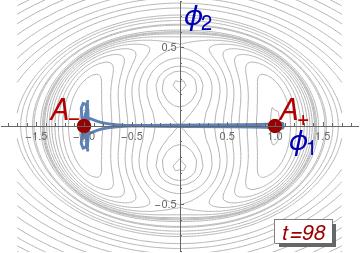}  }
\caption{\small  Evolution of multi-kink orbit (\ref{conca2}) for a $K^{(q,\lambda)}(v_0)$-$K^{(-q,-\lambda)}(-v_0)$ scattering process with impact velocity $v_0=0.7$ and model parameters $\tau=2.0$ and $\beta=0.2$.}
\end{figure}
The transition between the two previous regimes involves the presence of resonant windows, see Figure 13.

\noindent -- (4) For certain ranges of velocities $v_0$ the fluctuations of the second field component $\phi_2$ can change the previous behavior. In these cases the kink and the antikink involved in (\ref{conca2}) approach each other, collide and bounce back although vibrational eigenmodes are excited and radiation is emitted. As a whole, this type of events represents an inelastic kink-antikink reflection, similar to the scattering processes introduced in the first point (1). They can be characterized as
\begin{equation}
K^{(q,\lambda)}(v_0) \cup K^{(-q,-\lambda)}(-v_0) \rightarrow K^{(q,\lambda)}(-v_f) \cup K^{(-q,-\lambda)}(v_f)  + \nu \label{pheno4}
\end{equation}
where $v_f<v_0$ due of the transfer of kinetic energy to the vibrational modes and radiation emission.

\noindent -- (5) On other occasions the second field component fluctuations induce a novel scattering process. The kink and the antikink $K^{(q,\lambda)}(x)$ and $K^{(-q,-\lambda)}(x)$ collide, exchange its $\lambda$-charge and bounce back, establishing the following type of events
\begin{equation}
K^{(q,\lambda)}(v_0) \cup K^{(-q,-\lambda)}(-v_0) \rightarrow K^{(q,-\lambda)}(-v_f) \cup K^{(-q,\lambda)}(v_f)  + \nu \hspace{0.5cm} . \label{pheno1}
\end{equation}
Again, a part of the kinetic energy is employed to excite kink vibrational modes and to radiate, see Figure 13.

\subsection{${K}^{(q,\lambda)}(\overline{x})$-${K}^{(-q,\lambda)}(\overline{x})$ scattering processes }

Finally, in this subsection the study of the scattering processes between a Type II extended particle and its own antiparticle is addressed. In this case, the initial configuration, whose evolution must be analyzed, is given by the concatenation 
\begin{equation}
K^{(q,\lambda)}(x-x_0,t;v_0) \cup K^{(-q,\lambda)}(x+x_0,t;-v_0) \label{conca37}
\end{equation}
where $\lambda=\pm 1$ and $x_0$ is assumed to be large enough such that (\ref{conca37}) can be considered as a smooth function. This configuration has been depicted in Figure 6 together with its multi-kink orbit. The two field components of the initial configuration (\ref{conca37}) are now symmetric with respect to the spatial reflection $\pi_x$, so that scattering process of this type must preserve this symmetry. As before, the creation of Type I-Type II particle pairs is forbidden and now the emerging lumps must carry the same $\lambda$-charge. The kink scattering results are graphically represented in Figure 17 for three illustrative cases of our model: (a) $\tau=1.2$, $\beta=0.4$; (b) $\tau=2.0$, $\beta=0.2$ and (c) $\tau=3.0$, $\beta=0.2$. 

\begin{figure}[h]
	\centerline{\begin{tabular}{c}
			\includegraphics[height=2.cm]{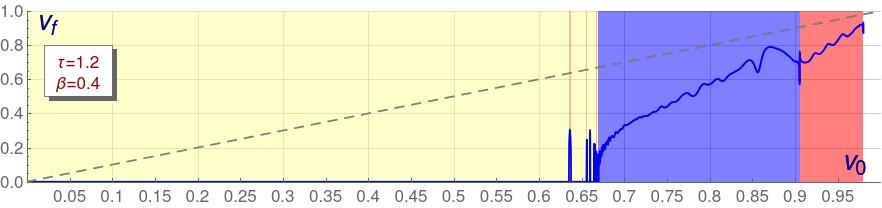}\\
			\includegraphics[height=2.cm]{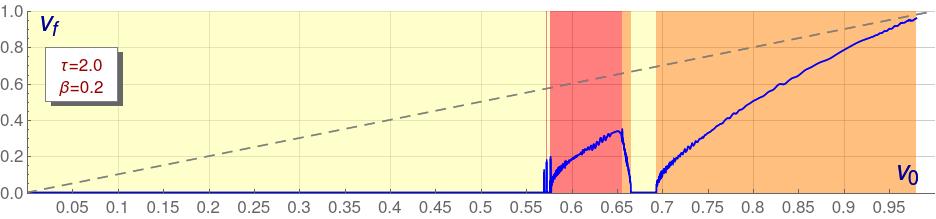} \\
		 	\includegraphics[height=2.cm]{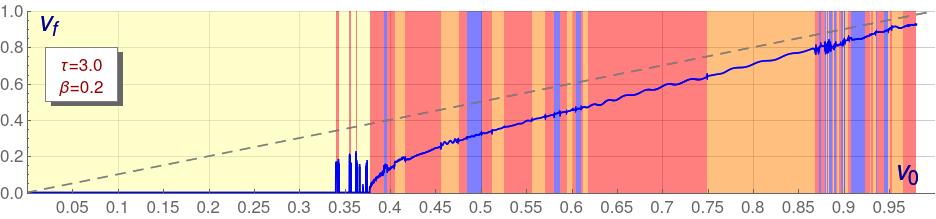}
	 	\end{tabular} \\ \hspace{0.3cm}
		\begin{tabular}{c} \includegraphics[height=2.cm]{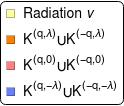}\end{tabular}}
	\caption{\small Graphical representation of the final velocity $v_f$ of the kink and the antikink as a function of the initial velocity $v_0$ for the ${K}^{(q,\lambda)}( \overline{x})$-${K}^{(-q,- \lambda)}( \overline{x})$ scattering processes in the cases $\tau=1.2$, $\beta=0.4$ (left, top), $\tau=2.0$, $\beta=0.2$ (left, middle) and $\tau=3.0$, $\beta=0.2$ (left, bottom). For the sake of comparison a dashed straight line has been added in these figures representing the final velocity in elastic collisions.  A color code distinguishing the possible events has been attached (right).}
\end{figure}
	
The general pattern revealed in Figure 17 is briefly described as follows: for all the three cases there is an annihilation regime for low initial velocities, as expected, since, in this case, the kink/antikink pair forms a bion, which periodically radiates energy. At the end of this regime the resonance windows are present. Here, the Type II particle and its antiparticle approach each other, collide and transmute into a Type I particle/antiparticle pair after the impact. These new lumps approach each other again, collide and bounce back several times before escaping by means of the resonant energy transfer mechanism. The final process can be represented as
\begin{equation}
K^{(q,\lambda)}(v_0) \cup K^{(-q,\lambda)}(-v_0) \rightarrow K^{(q,0)}(-v_f) \cup K^{(-q,0)}(v_f)  + \nu \hspace{0.5cm} . \label{pheno81}
\end{equation} 
The $\lambda$-charge vanishes for the two emerging lumps in this type of events. The details of the following kink scattering regimes are different for every case, as we can see in Figure 17:

\begin{enumerate}
\item For the first case with coupling constants $\tau=1.2$ and $\beta=0.4$, the collision of the original Type II particle/antiparticle pair changes the $\lambda$-charge of the colliding particles for initial velocities $v_0\in [0.67,0.905]$. Therefore, the emerging Type II particles carry opposite $\lambda$-charge than the original lumps, characterizing the scattering event given by
\begin{equation}
K^{(q,\lambda)}(v_0) \cup K^{(-q,\lambda)}(-v_0) \rightarrow K^{(q,-\lambda)}(-v_f) \cup K^{(-q,-\lambda)}(v_f)  + \nu \hspace{0.5cm} . \label{pheno8}
\end{equation}
A kinetic energy loss is induced by radiation emission after the impact, so $v_f<v_0$. For these cases the original semi-elliptic orbit (\ref{ellipse}) defined on a given half-plane ($\phi_2>0$ or $\phi_2<0$), evolves to the $\pi_2$-reflected trajectory, which changes the internal half-plane where the kinks live. This phenomenon has been depicted in the Figure 18. Observe the sign change of the second scalar field after the kink collision.

\begin{figure}[h]
	\centerline{\includegraphics[height=3.cm]{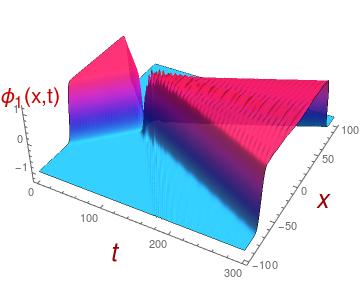} \hspace{1.5cm} \includegraphics[height=3.cm]{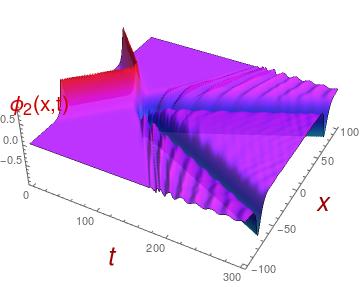} }
	\caption{\small  Evolution of the first and second scalar field components for a $K^{(q,\lambda)}(v_0)$-$K^{(-q,\lambda)}(-v_0)$ scattering process with impact velocity $v_0=0.7$ and model parameters $\tau=1.2$ and $\beta=0.4$.}
\end{figure}

For initial velocities greater than approximately 0.905, the Type II particle/antiparticle pair is transmuted into a Type I particle/antiparticle pair after the kink impact, whose constituents move away, giving rise to events of the type (\ref{pheno81}).

\item For the case $\tau=2.0$ and $\beta=0.2$ the situation is also remarkable. The previously mentioned resonance velocity windows precede the interval $[0.577,0.666]$ where two scattered kinks are found after the collision. If the impact speed is less than approximately $0.656$ the emerging particles are determined by the transmutation process (\ref{pheno81}), generating a Type I particle/antiparticle pair. Otherwise, an inelastic kink reflection of the form
\begin{equation}
K^{(q,\lambda)}(v_0) \cup K^{(-q,\lambda)}(-v_0) \rightarrow K^{(q,\lambda)}(-v_f) \cup K^{(-q,\lambda)}(v_f)  + \nu  \label{pheno82}
\end{equation}
takes place. This type of events is also found for values $v_0>0.693$. The most surprising aspect of the kink scattering displayed in Figure 17 (middle) is that a new annihilation regime arises for the interval $v_0\in [0.666,0.693]$. For these initial velocities the energy employed to excite the $\phi_2$-fluctuations is so large than there is no kinetic energy left to let the kinks to escape.  

\item Finally, if the model parameters are given by the values $\tau=3.0$ and $\beta=0.2$ a complex sequence of bands is found, see Figure 17 (bottom). Here, all the previous types of scattering events are interlaced.
\end{enumerate}

\section{Conclusions and further comments}

In this work the scattering between the two types of extended particles defined by the kink solutions of a two-component scalar field theory model has been analyzed. Type I particles carry topological charge $q=\pm 1$ whereas Type II particles are characterized by the charge pair $(q,\lambda)$ with $q,\lambda=\pm 1$. The symmetric scattering channels provided by the collisions between two Type II particles have been investigated. Within this category two types of scattering events can be distinguished depending on the $\lambda$-charge carried by the colliding lumps. If the colliding Type II particles have different $\lambda$-charge processes such as kink annihilation, transmutation of the original particles into two Type I particles, kink reflection or $\lambda$-charge exchange are possible. If the colliding Type II particles carry the same $\lambda$-charge then processes such as kink annihilation, transmutation, $\lambda$-charge reversing and kink reflection are found. The asymmetric scattering channels given by collisions between Type I and Type II particles are much more complex giving rise to ten possible different scattering events, as explained in Section 3.1. The presence of all the previous scattering channels depends crucially on the initial collision velocity. For every model characterized by the values of the coupling constants we find that the previous regimes are distributed in collision velocity bands where the previously mentioned scattering events are interlaced. This complexity is a characteristic mainly introduced by the increase of the number of fields in the model. 

This work opens new prospects for future research. From our point of view, it would be interesting to study the collisions between some non-topological kinks which arise in two-component scalar field theory models. The non-topological nature of these solutions could bring us new scattering events which have not been identified in this work.

\section*{Acknowledgments}

The author acknowledges the Spanish Ministerio de Econom\'{\i}a y Competitividad for financial support under grant MTM2014-57129-C2-1-P. He is also grateful to the Junta de Castilla y Le\'on for financial help under grant VA057U16. This research has made use of the high performance computing resources of the Castilla y Le\'on Supercomputing Center (SCAYLE, www.scayle.es), financed by the European Regional Development Fund (ERDF)

\end{document}